\documentclass[twocolumn,showpacs,amsmath,amssymb]{revtex4-1}
\usepackage{graphicx}
\usepackage{dcolumn}
\usepackage{bm}
\usepackage{amsmath}
\usepackage{amsfonts}
\usepackage{color}
\begin{document}

\title{Can Dark Matter explain the Braking Index of Neutron Stars?}
\author{Chris {\sc Kouvaris}$^1$, M. \'Angeles {\sc P\'erez-Garc\'ia}$^2$~\footnote{mperezga@usal.es}}
\affiliation{$^1$ $\text{CP}^3$-Origins, University of Southern Denmark, Campusvej 55, Odense 5230, Denmark\\ 
$^2$ Department of Fundamental Physics and IUFFyM, \\University of Salamanca, 
Plaza de la Merced s/n 37008 Salamanca, Spain}
\date{today}

\date{\today}
\begin{abstract}
We explore a new mechanism of slowing down the rotation of neutron stars via accretion of millicharged dark matter. 
 We find that this mechanism yields  pulsar braking indices that can be substantially smaller than the standard $n\sim 3$ of the magnetic dipole radiation model for millicharged  dark matter particles that are not excluded by existing experimental constraints thus accommodating existing observations. \\[.1cm] {\footnotesize \it Preprint: CP$^3$-Origins-2014-002
  DNRF90 \& DIAS-2014-2.}
\end{abstract}

\pacs{95.35.+d 95.30.Cq}
\maketitle 
Pulsar spin down has been studied since its discovery but yet to date, there is not a definite clear understanding for the handful of existing data values for rotation-powered pulsars \cite{living}. The braking index $n$ as defined from a spin down law $\dot \Omega \sim \Omega^n$,  where $\Omega$ is the rotational angular velocity, is consistently smaller than the $n=3$ value predicted for energy loss via pure dipole radiation. In order to explain this discrepancy other mechanisms have been invoked  including relativistic particle flow, and inclination angles or reconnection in the magnetosphere (see  e.g. the discussion in~\cite{yue} and~\cite{Contopoulos:2006mu}). Apart from short lived glitches  or eventual transitions to a more exotic quark star \cite{glendbook}, the rate of rotation of a neutron star (NS) drops steadily as a function of time. The deceleration of the rotation is due to torques that are applied on the star impeding its motion.
There are basically two components for the torque: an ``orthogonal'' one,  that maximizes  when the magnetic moment of the NS is perpendicular to the rotation axis of the star, and an ``aligned'' component, maximized when  the rotation and the magnetic axis of the star are aligned. In the ``orthogonal'' case, angular momentum is lost through emission of magnetic dipole radiation. In the ``aligned'' case, the torque is produced by the electric current created by escaping charged particles (mainly electrons and protons) that follow the open field lines of the NS magnetosphere. 
The rotational kinetic energy loss rate for both mechanisms present is~\cite{Conto05a}
\begin{equation}
L=L_{\rm orth} \sin^2 \theta + L_{\rm align} \cos^2 \theta, 
\label{Ltotal}
\end{equation}
where $\theta$ is the angle between the magnetic and the rotational axis. If we take $M$ and $R$ as the mass and radius of the NS respectively, the two  components can be written as 
\begin{equation}
L_{\rm orth} = \frac{B_0^2\Omega^4 R^6}{4c^3},\ \,\,\,\quad
L_{\rm align} = \frac{B_0\Omega \Omega_F R^3 I}{2c^2},
\label{lrate}
\end{equation}
\noindent
where $B_0$ is the magnetic field strength on the polar caps of the star, and $\Omega_F=\Omega-\Omega_{\rm death}$ ($\Omega_{\rm death}$ being the angular velocity below which the pulsar emission dies). 
The current $I$ is 
approximately equal to $I_{\rm GJ}$~\cite{Conto05b}, where $I_{\rm GJ}= \pi R_C^2 \rho_{\rm GJ}c$ is the current quoted in the pioneering work of Goldreich and Julian \cite{GJ}, representing the emission of relativistic  charged particles with
 charge density $\rho_{GJ}$, due to  a large difference in the electrostatic potential, from the NS cap regions of a surface $\sim  \pi R_C^2$.  $\rho_{\rm GJ}$ is the electron density which shields the electric field and
yields the only stable static solution for a pulsar magnetosphere (excluding centrifugal and gravitational forces).
At the poles of the NS, it was estimated to be~\cite{GJ}
\begin{equation}\rho_{\rm GJ}\simeq  7 \times 10^{-2}e \left(\frac{B_0}{10^{12}\,G}\right) \left(\frac{ s}{P} \right)~ \text{cm}^{-3},
\end{equation}  where $P$ is the period of the pulsar.

The field lines that do not close within the so-called light cylinder of the NS (defined as a cylinder of radius  $R_{\rm lc}=c/\Omega$), are the ones starting from the cap regions with an  angle  given by  $\theta_0$ where $\theta_0 \simeq \rm \arcsin \left( R/R_{\rm lc} \right)^{1/2}$. For a typical NS,  $R_{\rm C}=R\sin\theta_0\simeq 100\, \rm m$. Magnetic field lines starting outside of the caps return back to the star and therefore particles trapped in these lines do not contribute to a net current.  
If we use the definition of $I_{\rm GJ}$, $R_{\rm C}$, and $\theta_0$, we obtain \cite{Conto05b}, 
\begin{equation}
I=I_{\rm GJ}= \frac{B_0R^3 \Omega^2}{2c} \simeq 1.4 \times 10^{30} e \left(\frac{B_0}{10^{12}\,G}\right) \left(\frac{ s}{P} \right)^2~\text{s}^{-1}.
\label{IGJ}
\end{equation}
However as we shall argue, if  dark matter (DM) is in the form of millicharged  particles (MC) (as for example~\cite{goldberg},~\cite{Kouvaris:2013gya}), accretion of DM onto the NS can lead to an excess of electric charge that due to overall electric neutrality must be expelled from the caps, providing  this way an extra component of current.
On generic grounds, let us assume  that the total current has two components, $I=I_{\rm GJ} + I_{\rm DM}$, where $I_{\rm DM}$ represents the extra current due to external accretion from the MCDM galactic distribution where the NS may be located~\cite{SanchezSalcedo:2010ev}. 

At this point let us recall that the rotational kinetic energy of a nearly spherical NS (upon assuming uniform density for most of the star in the core central region) is $E \simeq \frac{1}{2}{\cal I} \Omega^2$ where at lowest order  ${\cal I}=\frac{2}{5}MR^2$ is the moment of inertia of the star~\cite{LS2005}. The rotational kinetic energy loss rate is $\dot{E}={\cal I} \Omega \dot{\Omega}=\frac{2}{5} MR^2\Omega \dot{\Omega}=-L$. Using the above along with Eqs.~(\ref{Ltotal}) and (\ref{IGJ}), we obtain
\begin{equation}
\small
\dot{\Omega}=-\frac{5B_0^2\Omega^3R^4}{8Mc^3} \left [ \sin^2 \theta + \left (1-\frac{\Omega_{\rm death}} {\Omega} \right) \left ( 1+\frac{I_{\rm DM}}{I_{\rm GJ}} \right ) \cos^2\theta \right ]. \label{dotO}
\end{equation}

As it can be seen, $\dot{\Omega}$ depends on $B_0$. In fact the rate of deceleration of the rotation is usually used to deduce the magnitude of the magnetic field. Instead, the braking index $n=\ddot{\Omega}\Omega/\dot{\Omega}^2$, is independent of the magnetic field strength. Using the definition of $n$ and Eq.~(\ref{dotO}) we find 
\begin{equation}
n=3-\frac{2 \left(1-\frac{\Omega_{\rm death}}{\Omega} \right ) \lambda - \frac{\Omega_{\rm death}}{\Omega} \left ( 1+ \lambda \right )}{\sin^2\theta +\left (1-\frac{\Omega_{\rm death}}{\Omega} \right )(1+\lambda)\cos^2\theta}\cos^2\theta,
\end{equation}
where $\lambda=\frac{I_{\rm DM}}{I_{\rm GJ}}$. If we assume that the pulsar is younger than 10 to 100 million years, we can safely assume $\frac{\Omega_{\rm death}}{\Omega}<<1$. In that case $n$ takes the simple form
\begin{equation}
n=3-\frac{2\lambda \cos^2\theta}{1+\lambda \cos^2\theta}.
\end{equation}
As we can see, for $\lambda=0$ (no extra current), we recover $n=3$ for any angle $\theta$, which is the well known prediction for the aligned rotator as well as the magnetic dipole radiation. $\lambda$ can take positive (negative) values depending on the relative charge sign of the extra $I_{\rm DM}$ current with respect to $I_{\rm GJ}$. However, one can see that for $|\lambda|\sim {\cal O}(1)$, the braking index can be either smaller ($\lambda>0$) or  larger ($\lambda<0$) than 3. There are not many stars where the braking index has been estimated \cite{yue}, seeming to range from $n\sim$1.4 to 2.91. If $I_{\rm DM}$ has the same sign with $I_{\rm GJ}$ and it is roughly of the same order of magnitude, braking indices $n\sim$2 or less are possible. To examine this possibility we need to find if indeed accreted DM in the form of MC particles can accommodate such a current.

The capture rate of MCDM strongly depends  on the the mass $m$ and electric charge ($q=\epsilon e$, in units of the electron charge $e$) of the particle constituent. For sufficiently small electric charge and/or large mass for the particle,  the flux of DM particles crossing the surface of the star is determined by the gravitational attraction of the star. However, if the electric charge is sufficiently large, then an enhanced number of MCDM particles may  follow the magnetic lines (which are electrically equipotential) in the magnetosphere within the light cylinder and cross the surface of the star. On general grounds, we will show that the latter case may lead to a much higher accretion rate. 

In this case,  if the Larmor radius $r_L=\frac{mv_{\perp}}{\epsilon e B}$, is  much smaller than the 
 curvature radius of the star's magnetic field defined as $R_{\rm curv}(r, \gamma)=\left |\frac{1}{B}\frac{dB}{dr} \right |^{-1}$,  the particle follows in a helical fashion the magnetic field line  eventually crossing the star and thus getting trapped.  One should keep in mind that $R_{\rm curv}(r, \gamma)$ is a function of both the radial distance $r$ and the latitude $\gamma=\frac{\pi}{2}-\theta$. 
$v_{\perp}$ is  particle's velocity perpendicular to the $B$ field line. The B field strength corresponds to the poloidal component within the light cylinder given in~ \cite{GJ} as 
\begin{equation}
B_p=\frac{1}{2}\frac{R^3}{R_{\rm lc} r^2}B_0 \chi, \label{pol}
\end{equation}
 where $\chi \sim {\cal O}(1)$ is a coefficient.  For the sake of the estimate and in what follows we will consider the equatorial latitude, $\gamma\simeq 0$, as an upper limit on the MCDM capture rate.
The condition for capture is $r_L<<R_{\rm curv}$. Using the explicit value of the magnetic poloidal field strength of Eq.~(\ref{pol}), we find that this is satisfied as long as $r_L<<r/2$. $r_L$ depends on the velocity of the MC particle. Far away from the NS, assuming a thermal DM distribution as e.g. in the solar neighbourhood, the particle's velocity is $v_0\sim 230 \,\text{km}/\text{s}$, but as the particle approaches the star, it is gravitationally boosted to $v=\sqrt{v_0^2+2GM/r}$.  Inserting this expression and requiring $r_L<<r/2$ leads to the condition
 \begin{equation}
 r<<r_0=\frac{GM}{v_0^2}(\sqrt{1+\Delta^2}-1),
 \end{equation}
 where $\Delta=\epsilon e v_0 \Omega R^3 B_0 \chi/(4GMm)$, $M$ being the mass of the star. 
%

The above condition means that for $r<<r_0$ one cannot ignore the helical motion trapping effect in the magnetosphere. One should emphasize that satisfying this relation does not necessarily mean that all particles will follow the magnetic lines and hit the star. Whether a particle can be trapped by following the magnetic lines or not depends also on the pitch angle between particle's velocity and the magnetic field. However, if the condition is satisfied,  a substantial fraction of the particles entering the area will definitely be trapped. The exact fraction  will depend on the detailed kinematics in the accretion process arising from the DM distribution.

A rough estimate of the accreted number of particles can be obtained by simply finding the flux of particles entering through the surface that defines the minimum volume between the light cylinder and a sphere of radius $r_0$. This can be evaluated approximately by considering the light cylinder as a sphere with radius $R_{\rm lc}=c/\Omega$. In this case, the smaller between $r_0$ and $R_{\rm lc}$ defines the sphere which if millicharged particles cross its surface, a significant fraction of them will be captured. 

Let us calculate the flux of particles 
  entering a sphere of radius $R_*$.
Assuming a typical Maxwell-Boltzmann distribution for the velocities of DM particles, the flux of particles crossing the surface with velocity between $v$ and $v+dv$ and with angle with respect to the normal $\phi$ and $\phi + d \phi$ is
  \begin{equation}
  dF=n_0 \left ( \frac{3}{2\pi v_0^2} \right )^{3/2} \pi v^3 \exp \left ( \frac{-3v^2}{2v_0^2} \right ) d\cos^2\phi dv,
  \end{equation}
 where $n_0$ is the  number density of DM particles, in the neighborhood of the star. The total flux (number of particles per time) crossing the surface of a sphere with radius $r$ can be written in terms of two new variables: (non-relativistic) energy per mass $E=\frac{1}{2}v^2$ and angular momentum per mass $J=vr \sin \phi$ as
\small
  \begin{equation}
  dF_{\rm total}=4\pi r^2dF=n_0 \left ( \frac{3}{2\pi v_0^2} \right )^{3/2} \exp \left ( \frac{-3v^2}{2v_0^2} \right ) 4\pi^2dEdJ^2. 
\label{DFT}
  \end{equation} 
\normalsize
  The perihelion of the orbit of a particle around a mass $M$ in Newtonian dynamics  given in terms of $E$ and $J$ is $r_{\rm per}=a/(1+\sqrt{1+2ab})$ where $a=J^2/GM$ and $b=E/GM$. We demand  the perihelion to be smaller than the radius of the sphere under consideration i.e. $r_{\rm per}<R_*$. From this we determine the range of $J^2<2GMR_*(1+ER/GM)$. Integrating Eq.~(\ref{DFT}) first over $J^2$ from zero to the maximum value, and then over $E$ from zero to infinity, we get 
  \begin{equation}
  F_{\rm total}=n_0\frac{8\pi^2}{3v_0}\left (\frac{3}{2\pi} \right )^{3/2} GMR_* \left (1+\frac{v_0^2}{3}\frac{R_*}{GM} \right).
  \end{equation} 
\normalsize
Let us discuss the different conditions arising in the capture process. In the case where $r_0>>R$ (case I), the total flux of particles captured is given approximately by the above equation with $R_*\simeq \rm min(r_0, R_{lc})$. For a rotation period $P=1$ s ($P=0.1$ s) we found that if $\epsilon \gtrsim 10^{-5.5}m$ ($\epsilon \gtrsim 10^{-7}m$) with $m$ being measured in GeV, $R_{\rm lc}$ is always the minimum and the flux  for a typical NS of $M=1.4 M_{\odot}$ and $R=10$ km is
\small
\begin{equation}
 F_{\rm I}\simeq  1.0 \times 10^{29} \left (\frac{\rho_{\rm DM}}{0.3 \, \text{GeV}\text{cm}^{-3}} \right )\left (\frac{1\rm GeV}{m} \right )\left (\frac{P}{\rm s} \right )~\text{s}^{-1}.  
\label{acc1}
  \end{equation}
\normalsize
If $r_0<<R$  (case II) one can completely ignore the magnetic field and the accretion  proceeds as in the case of a neutral particle.  
The capture rate of such particles is ~\cite{Kouvaris:2011fi}
  \begin{equation}
F_{\rm II}=\frac{8\pi^2}{3} \frac{\rho_{\rm DM}}{m} \left(\frac{3}{2 \pi v^2} \right )^{3/2}\frac{GMR}{1-\frac{2GM}{R}}v^2(1-e^{-3E_0/v^2})f, \label{acc12}
  \end{equation}
where $E_0$ is the maximum kinetic energy per mass that can lead to capture  and $f$ is a capture efficiency factor estimated in~\cite{Kouvaris:2007ay}. The flux for a typical NS is then
\begin{equation}
F_{\rm II}= 4.2 \times 10^{26} \left (\frac{\rho_{\rm DM}}{0.3\,\text{GeV}\text{cm}^{-3}} \right )\left (\frac{1\rm GeV}{m} \right )f~\text{s}^{-1}. 
\label{acc2}
\end{equation}
The accretion of MCDM particles onto the NS will lead at some point to an equilibrium between incoming and outgoing electric charge. If particles with electric charge $\epsilon e$ are accumulated in the NS with a rate given in Eq.~(\ref{acc1}), and due to the fact that the NS is an excellent electric conductor, the charge will be distributed at the surface. It is easy to see that for an electron or proton  even if we ignore the fact that there is a magnetosphere, and therefore an electric potential on the surface of the star, the gravitational force becomes smaller than the Coulomb repulsive force once
\begin{equation}
\frac{GMm}{R}<\frac{N \epsilon \alpha \hbar c }{R}, \label{n1}
\end{equation}
where $N$ is the number of MC particles trapped in the star, and $\alpha$ is the fine structure constant. For example, for a proton with mass $m_p$, once $ N \gtrsim 1.46 \times 10^{21} \epsilon^{-1} m_p$  ($m_p$ measured in GeV), the Coulomb force is already larger than the gravitational force, leading to expulsion of charge from the star. For electrons, this will happen roughly three orders of magnitude lower than the protons. Once equilibrium between accretion and expulsion is established, then the electric charge expelled by the star  should balance that of the accretion.

Some comments are in order here. One might naively expect that the accumulating electric charge onto the NS could stop the further accretion of DM particles due to Coulomb repulsion. However, one can trivially check that in order for the repulsive Coulomb force to compete with the gravitational force in the case of a MC particle, the corresponding Eq.~(\ref{n1}) should now give  $ N \gtrsim 1.4 \times 10^{21}\epsilon^{-2} m$, where $m$ is the mass of the MC particle in GeV. Electrons or protons (depending on the charge of DM) are getting expelled by the star before accretion can be stopped as long as $\epsilon (m_p/m)<<1$, which  is satisfied for all cases of our interest.

Under these conditions, once equilibrium between accretion and expulsion of electric charge has been established,  the expelled charge will be in the form of either electrons or protons (depending on the charge of the accreted MC particles) because as long as $\epsilon (m_{p,e}/m)<<1$ is satisfied ($m_{p,e}$ being the mass of the proton or the electron), protons or electrons are easier to expel than MC particles. 

Finally we should comment on our approximation on the addition of $I_{\rm GJ}$ and $I_{\rm DM}$ components in order to get the total current. In principle, one has to solve the problem of the magnetosphere of the neutron star from scratch. In the absence of accretion of charged particles $\vec{E}\cdot  \vec{B}=0$. In addition, the potential of the cap is also determined by demanding overall neutrality on the emitting particles from the surface of the star to the magnetosphere. In our case, the presence of extra charge on the surface of the NS will make $\vec{E}\cdot  \vec{B}\neq 0$ and definitely there is not an overall neutrality since it must expel the excessive charge accumulated by the accreted DM. Despite all this, it is a good first approximation to add up the currents. First of all even in  non accreting NSs, $\vec{E}\cdot  \vec{B}$ is never zero in reality. In addition one can compare the total charge on the surface of the star to the charge we estimated earlier needed to start expelling protons or electrons from the caps. The charge on the surface of a NS is $(2/3)B\Omega R^3e^2/\alpha \sim 3\times 10^{29}e$ (for a typical $B=10^{12}$ G and $P=1~\rm s$). This number is much larger than the $1.4 \times 10^{21}\epsilon^{-2} m$ we derived earlier for expelling charges from the star.  
 Therefore, the extra charge (coming from DM) should be a small perturbation to the solution of the magnetosphere without MC particles. However, and this is a crucial point, the current $I_{\rm DM}$ can be a substantial fraction of $I_{\rm GJ}$. It is expected that the excessive charge (positive or negative) will follow exactly the pattern of expulsion from the cap of the star, thus justifying the addition of the two components.
 
In order to evaluate the two  cases i.e. electromagnetic accretion  (case I) and  gravitational accretion (case II) we calculate the conditions such that $\lambda=\frac{I_{\rm DM}}{I_{\rm GJ}}\sim 1$. Using  $I_{\rm DM} \simeq \epsilon e F_{\rm I}$  for case I we obtain the condition,
\small
\begin{equation}
\epsilon \left (\frac{\rm GeV}{m} \right ) \simeq 14\left (\frac{0.3\,\text{GeV}\text{cm}^{-3}}{\rho_{\rm DM}} \right ) \left (\frac{\rm s} {P}\right )^{3} \left(\frac{B_0}{10^{12}\,G}\right), 
\label{res1}
\end{equation}
\normalsize 
while for case II with $I_{\rm DM}\simeq  \epsilon e F_{\rm II}$ we get
\small
\begin{equation}
\epsilon \left (\frac{\rm GeV}{m} \right ) \simeq \frac{3.3 \times 10^3}{f} \left (\frac{0.3\,\text{GeV}\text{cm}^{-3}}{\rho_{\rm DM}} \right ) \left (\frac{\rm s} {P}\right )^{2} \left(\frac{B_0}{10^{12}\,G}\right).
\label{res2}
\end{equation}
\normalsize
At this point one should note that MCDM is already constrained by experimental data~\cite{lhc, Dimopoulos:1989hk,Davidson:1993sj,c1,c2,prinz,Davidson:2000hf,Dubovsky:2003yn,lang}. These bounds have been obtained by a variety of different methods: accelerator and collider experiments, invisible decay of orth-positronium, MC particle searches, Lamb shift, Big Bang Nucleosynthesis, plasmon decay in white dwarfs, dark matter searches and Supernova 1987A. All the above constrain the $\epsilon-m$ phase space of MC particles. The reader can see Fig. 1 of~\cite{Davidson:2000hf} for a comprehensive  explanation of the different bounds. 
In addition there are cosmological bounds based on CMB, constraining the MC particle fraction with some underlying theoretical model dependence \cite{Dubovsky:2003yn,dolgov}. It is worth mentioning that additional forms of charged DM under the strangelet, Q-balls and other phenomenology may also arise in NS catastrophic events \cite{str} although at an expected much lower flux.

One can check  that MCDM with mass $m\sim 1$ MeV (1 GeV), $\epsilon \sim 10^{-4}$ (0.1) and with a density $\rho_{\rm DM}$ of a  few~ $\rm GeV/\rm cm^3$ satisfies
Eq.~(\ref{res1}) while the same time does not violate experimental bounds.  It has been suggested~\cite{Chuzhoy:2008zy} that  magnetic galactic fields can prevent  MC particles from the halo to enter the galactic disk and those initially trapped in the disc can be accelerated and injected out of the disc within 0.1 to 1 billion years, if MC particles fall within $100\epsilon^2 \lesssim m \lesssim 10^8 \epsilon$ TeV. However, even if this is true, the two examples we gave above (i.e. the 1 MeV and the 1 GeV particles) lie exactly at the left margin of the inequality and therefore it is not clear if the constraint is applicaple. In addition, there is a large number of NS lying out of the galactic plane where the above constraint is irrelevant.

There are two remarks we would like to make here. The first one is that $P>1$ s, DM  densities larger than $\sim 1~ \rm GeV/ \rm cm^3$ and/or smaller values of $B$ (i.e. $B<10^{12}$ G)   allow more phase space in terms of $\epsilon$ and $m$ in order to satisfy Eqs.~(\ref{res1}), (\ref{res2}). Pulsars can potentially have periods up to a few seconds and DM densities close to the center of the Galaxy can be substantially larger than that around the Earth. As for the magnetic field, we should emphasize that usually in a NS its value is estimated 
indirectly by observing the change in the period of the pulsar, $dP/dt$, and by applying the main mechanisms of angular momentum loss.  However, for $B<10^{12}$ G, Eqs.~(\ref{res1}), (\ref{res2}) are easier satisfied and  $dP/dt$ can still be the observed one despite the magnetic field being smaller. The second remark has to do with  the sign of accreted charge and the change in the braking index of the NS. If $\Omega \cdot B>0$ at the poles, accretion of negative (positive) MCDM will reduce (increase)  the braking index of the star wth respect to the value three of the ``aligned rotator". Obviously it will be the opposite in case the magnetic field and the rotation axis are antiparallel. 

So far we discussed  the accretion of DM particles with definite charge into NSs. If  DM consists of MC particles and it is overall electrically neutral, there are two possibilities: i) 
 MC particles appear  as positive or negative equal  charge (as an absolute value) and mass. In this case the scenario we present in this paper is not valid simply because the net accumulated charge is zero. In principle fluctuations in the accretion can induce a net current, but we found that unless the star is immersed in a DM environment of extremely high density, there is no effect. ii) DM  is overall electrically neutral but it consists of two components of positive and negative MC particles with different charge and/or mass in such a way that  
 accretion onto the star is different for the positive and the negative MC particles. This is not difficult to achieve since, as we pointed out, accretion rates depend strongly on the $\epsilon/m$ ratio for the particle species. If for example the negative MC particles are accumulated onto the star via type I accretion (Eq.~(\ref{acc1})) and the positive ones via type II (Eq.~(\ref{acc2})), the NS will dominantly accrete the negative component and therefore the braking index will become smaller than 3 (if $\Omega \cdot B>0$ at the poles).
In addition, even within the same regime (I or II), accretion rates can still be different between positive and negative components. For example by inspection of  Eq.~(\ref{acc12}), one can see that different particle masses can lead to different accretion rates. 

In this work we present the possibility that accretion of a  two-component millicharged dark matter onto a neutron star even at relatively low dark matter densities can significantly change the braking index of a pulsar. For MC particles that are still evasive of experimental constraints, accretion onto neutron stars can lead to expulsion of extra electric charge from the poles of the star, which consequently can impede further the rotation of the star yielding braking indices consistent with the ones observed.

CK would like to thank I. Contopoulos for very useful discussions. C.K. is supported by the Danish National Research Foundation, Grant No. DNRF90. M.A.P.G. would like to thank the newCOMPSTAR, and the MULTIDARK projects, as well as CP$^3$-Origins where part of this work was developed. M.A.P.G. is supported by the
Spanish MICINN project  FIS2012-30926. 
   

\begin{thebibliography}{99}	

\bibitem{living} M. A. Livingstone et al., Astrophysics and Space Sciences {\bf 308} (2007) 317.
\bibitem{yue}Y. L. Yue, R. X. Xu, W. W. Zhu, Adv. in Space Research {\bf 40} (2007) 1491.

\bibitem{Contopoulos:2006mu} 
  I.~Contopoulos,
  [astro-ph/0610156].


\bibitem{glendbook} N. K. Glendenning, N.K., {\it Compact stars}, Ed. Springer-Verlag, New York ( 2000)

\bibitem{Conto05a} A. K. Harding, I.Contopoulos, and D. Kazanas, ApJ {\bf 525} L125 (1999)

\bibitem{Conto05b} I. Contopoulos and A. Spitkovsky, ApJ, {\bf 643}, 1139 (2006).

\bibitem{GJ}  P.~Goldreich and W.~H.~Julian, ApJ\  {\bf 157}, 869 (1969).  

\bibitem{goldberg}  H. Goldberg and L. J. Hall, Phys. Lett. B {\bf174} (1986) 151.

\bibitem{Kouvaris:2013gya} 
  C.~Kouvaris,
  Phys.\ Rev.\ D {\bf 88}, 015001 (2013)
  [arXiv:1304.7476 [hep-ph]].
  
\bibitem{SanchezSalcedo:2010ev} 
  F.~J.~Sanchez-Salcedo, E.~Martinez-Gomez and J.~Magana,
  JCAP {\bf 1002}, 031 (2010)
  [arXiv:1002.3145 [astro-ph.CO]].
  


\bibitem{LS2005} J. M. Lattimer and B. F. Schutz, ApJ {\bf 629} (2005) 979 .
 \bibitem{Kouvaris:2011fi}   C.~Kouvaris and P.~Tinyakov,  Phys.\ Rev.\ Lett.\  {\bf 107}, (2011) 091301.
\bibitem{Kouvaris:2007ay}  C.~Kouvaris,  Phys.\ Rev.\  D {\bf 77}, (2008)  023006.

\bibitem{lhc}J. Jaeckel, M. Jankowiak and M. Spannowsky, Phys.Dark Univ. 2 (2013) 111.

\bibitem{Dimopoulos:1989hk} 
  S.~Dimopoulos, D.~Eichler, R.~Esmailzadeh and G.~D.~Starkman,
  Phys.\ Rev.\ D {\bf 41}, 2388 (1990).
 

\bibitem{Davidson:1993sj} 
  S.~Davidson and M.~E.~Peskin,
  Phys.\ Rev.\ D {\bf 49}, 2114 (1994)
  [hep-ph/9310288].
  
\bibitem{c1}
R.~N.~Mohapatra and I.~Z.~Rothstein,
 Phys.\ Lett.\ B {\bf 247}, 593 (1990).

\bibitem{c2}
R.~N.~Mohapatra and S.~Nussinov,
 Int.\ J.\ Mod.\ Phys.\ A {\bf 7}, 3817 (1992).

\bibitem{prinz} A.A. Prinz et al., Phys. Rev. Lett. {\bf 81} (1998) 1175.

\bibitem{Davidson:2000hf} 
  S.~Davidson, S.~Hannestad and G.~Raffelt,
  JHEP {\bf 0005}, 003 (2000)
  [hep-ph/0001179].



\bibitem{Dubovsky:2003yn} 
  S.~L.~Dubovsky, D.~S.~Gorbunov and G.~I.~Rubtsov,
  JETP Lett.\  {\bf 79}, 1 (2004)
  [Pisma Zh.\ Eksp.\ Teor.\ Fiz.\  {\bf 79}, 3 (2004)]
  [hep-ph/0311189].

\bibitem{lang} P. Langacker and G. Steigman, Phys. Rev. D {\bf 84} (2011) 065040.

\bibitem{Burrage:2009yz} 
  C.~Burrage, J.~Jaeckel, J.~Redondo and A.~Ringwald,
  JCAP {\bf 0911}, 002 (2009)
  [arXiv:0909.0649 [astro-ph.CO]].

 

\bibitem{dolgov} A.D. Dolgov et al., Phys. Rev. D {\bf 88}(2013)  117701.
\bibitem{str} M. A. Perez-Garcia, F. Daigne, and J. Silk, ApJ {\bf 768} (2013) 145 [arXiv:1303.2697 [astro-ph.HE]]; M. A. Perez-Garcia, J. Silk and J. R. Stone, Phys. Rev. Lett. {\bf 105} (2010) 141101 [arXiv:1108.5206v1 [astro-ph.CO]].

\bibitem{Chuzhoy:2008zy} 
  L.~Chuzhoy and E.~W.~Kolb,
  JCAP {\bf 0907}, 014 (2009)
  [arXiv:0809.0436 [astro-ph]].
[arXiv:1303.2697 [astro-ph.HE]]

    \end{thebibliography}
  \end{document}